\documentclass[12pt, letterpaper]{article}
\usepackage[top=1in, bottom=1.5in, left=1in, right=1in]{geometry}
\usepackage[utf8]{inputenc}
\usepackage{booktabs}
\usepackage{natbib}
\usepackage[colorlinks=true,linkcolor=blue, citecolor=blue, linkcolor=red]{hyperref}
\usepackage{graphicx}
\usepackage{xcolor}
\usepackage{authblk}
\definecolor{green}{HTML}{228B22}

\title{Unveiling the Rich and Diverse Universe of Subsecond Astrophysics through LSST Star Trails}
\author{
\parbox{0.9\textwidth}{
\normalsize
David Thomas$^{1,2}$,
Steven M. Kahn$^{2,3}$,
Federica B. Bianco$^{4,5}$,
Željko Ivezić$^{6}$,
Claudia M. Raiteri$^{7}$,
Andrea Possenti$^{8}$,
John R. Peterson$^{9}$,
Colin J. Burke$^{10}$,
Robert D. Blum$^{11}$,
George H. Jacoby$^{12}$,
Steve B. Howell$^{13}$,
Grzegorz Madejski$^{2}$,
\textit{with the support of the LSST Transients and Variable Stars Collaboration
}}}
\date{November 2018}

\begin{document}

\maketitle

\begin{abstract}
We present a unique method that allows the LSST to scan the sky for stellar variability on short timescales.
%The method has operational and image processing components.
The operational component of the strategy requires LSST to take star trail images. The image processing component uses deep learning to sift for transient events on timescales down to 10 ms. We advocate for enabling this observing mode with LSST, as coupling this capability with the LSST's tremendous 319.5 m$^2$deg$^2$ etendue will produce the first wide area optical survey of the universe on these timescales. We explain how these data will advance both planned lines of investigation and enable new research in the areas of stellar flares, cataclysmic variables, active galactic nuclei, Kuiper Belt objects, gamma-ray bursts, and fast radio bursts.
\end{abstract}

\section{White Paper Information}
% The corresponding author is David Thomas (dthomas5@stanford.edu). 
% This work was developed partly within the Transients and Variable Stars Science Collaboration (TVS) and the author acknowledges the support of TVS in the preparation of this paper.

\begin{enumerate} 
\item {\bf Science Category:} We introduce a mechanism that enables the LSST to provide new data with higher time resolution that not only enhances existing investigations, but allows LSST to contribute to new science use cases that generally lie within the categories: \textit{Exploring the Transient Optical Sky}, \textit{Mapping the Milky Way}, and \textit{Taking an Inventory of the Solar System}.

\item {\bf Survey Type Category:} This strategy could be implemented as a \emph{Mini-survey}, or by inserting occasional star trail images into the main \textit{Wide-Fast-Deep} survey.

\item {\bf Observing Strategy Category:}
While different fields are conducive to different aspects of our method - for example, searching open clusters for flare stars - it is largely agnostic of where the telescope is pointed. Furthermore, our proposal can be trivially interleaved with the main LSST survey.

\end{enumerate}

\clearpage

\section{Scientific Motivation}

% \begin{footnotesize}
% {\it Describe the scientific justification for this white paper in the context
% of your field, as well as the importance to the general program of astronomy, 
% including the relevance over the next decade. 
% Describe other relevant data, and justify why LSST is the best facility for these observations.
% (Limit: 2 pages + 1 page for figures.)}
% \end{footnotesize}

Subsecond photometry provides particularly valuable insights for studies of compact 
objects and for studies of occultations, eclipses, and transits. In the case of compact objects, such as stellar mass black holes, white dwarfs and neutron stars, high-speed observations are required because their dynamical timescales range from milliseconds in black holes and neutron stars to seconds in white dwarfs. In the case of extrinsic variability, such as occultations and eclipses, increased time resolution typically corresponds to increased spatial resolution. Subsecond photometry to faint brightness levels can thus open up new parameter space in studies of a wide range of astrophysical phenomena including active galactic nuclei (AGN), stellar flares, Kuiper Belt objects (including their atmospheres), fast radio bursts (FRBs), gamma-ray bursts (GRBs), X-ray binaries, polars, symbiotic stars, cataclysmic variables, asteroid sizes, and might as well discover new sources of photometric variability (e.g., \citealt{D2011, KRS2014, Gandhi2016}). 

Unfortunately, subsecond photometry to faint brightness levels is technically challenging, especially if large sky coverage is required. Conventional optical telescopes rely on charge-coupled devices (CCDs) which typically take at least several seconds to read out. This readout time limits the time resolution they can achieve and precludes them from efficient high-time-resolution investigations. Alternative instrument technologies that can image optical bands at high speeds have fields of view that are typically a few arcminutes, or less than 1/1000th of the LSST field of view. We present here a method that allows the LSST to explore the subsecond universe at an unprecedented combination of 
depth and sky coverage. 

\begin{figure*}[htb]
\center
\includegraphics[width=1.00\columnwidth]{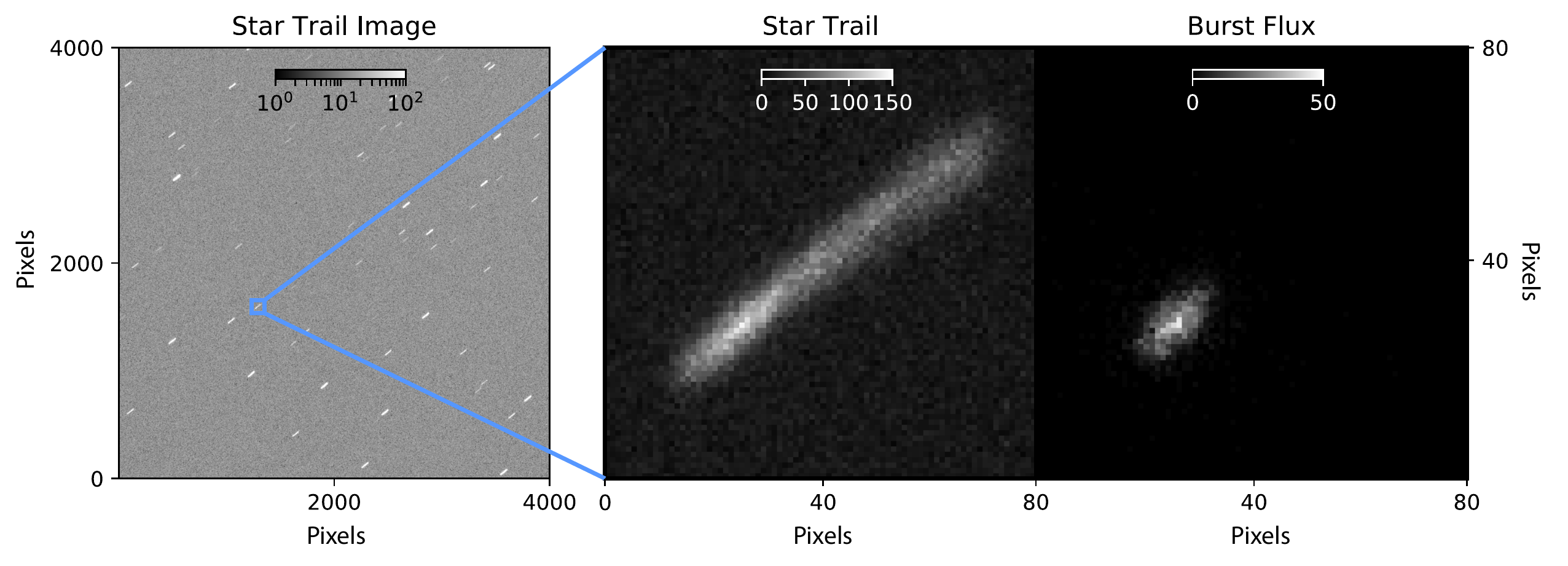}
\caption{A star trail image corresponding to a 1 second exposure on a single LSST CCD in the \textit{r} filter. The two zoom-ins on the right show a star trail exhibiting variability.}
\label{fig:trail}
% http://localhost:8888/notebooks/Code/Astronomy/exp7/MakeWhitePaperGraphics_20181120.ipynb
\end{figure*}

\begin{figure*}[htb!]
\center
\includegraphics[width=\columnwidth]{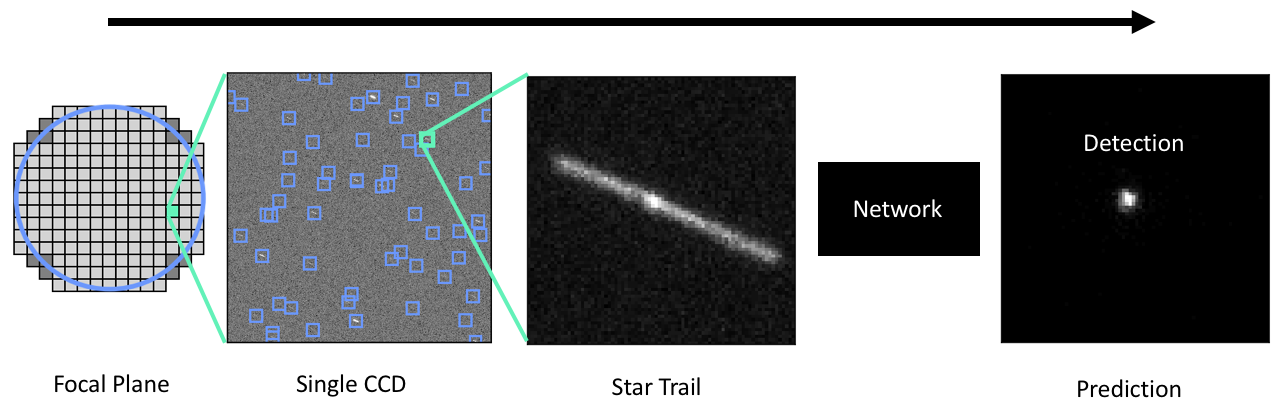}
\caption{The image processing pipeline. We extract the individual star trails from the image and classify them with a neural network.}
\label{fig:pipeline}
% http://localhost:8888/notebooks/Code/Astronomy/exp4/notebooks/MakeVariabilityGraphic_20180329-Copy1ForICMETalk.ipynb
\end{figure*}

\begin{figure*}[htb!]
\center
\includegraphics{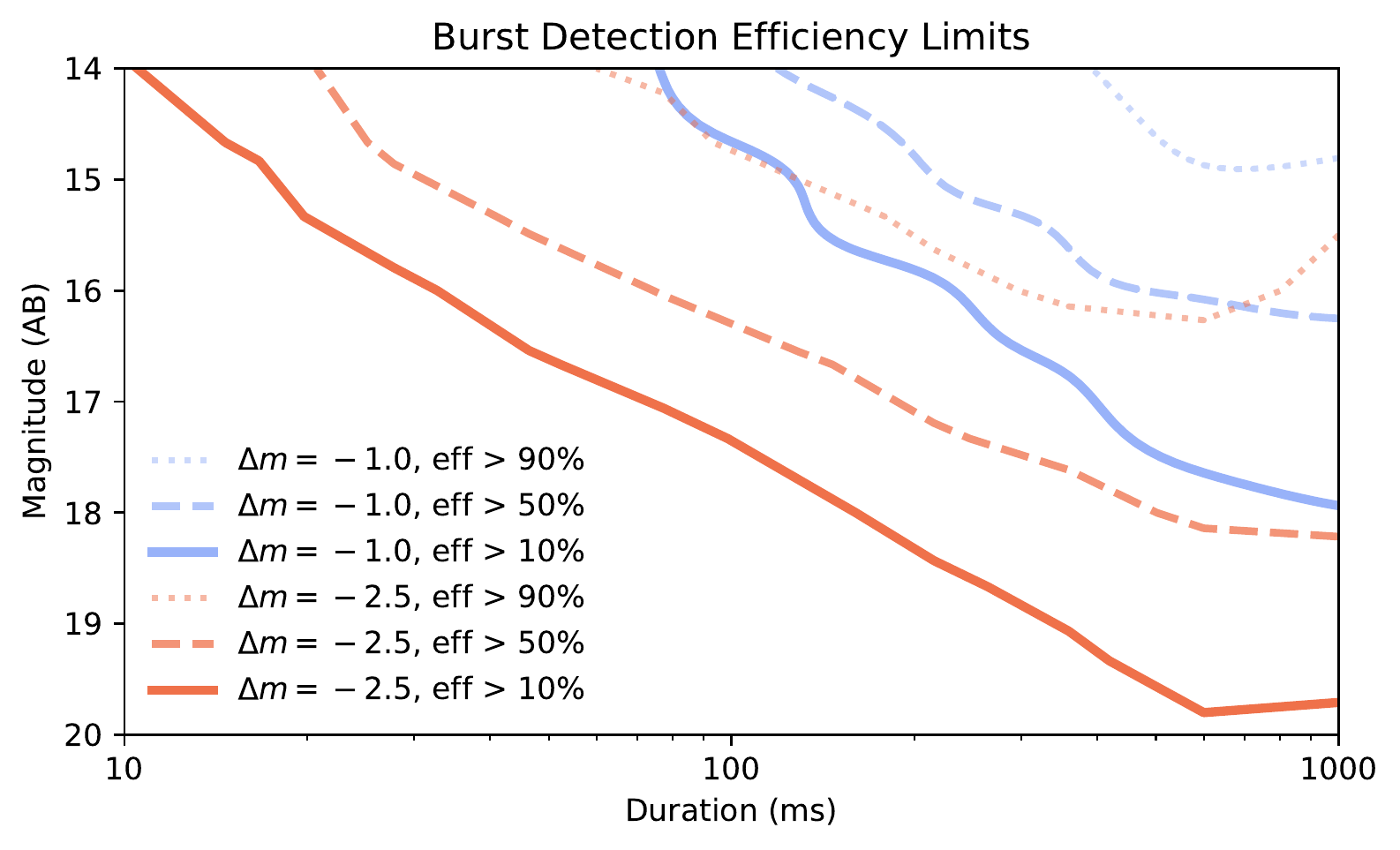}
\caption{Detection accuracy and performance limits for 15 second star trails.}
\label{fig:longlimit}
% http://localhost:8888/notebooks/Code/Astronomy/exp4/notebooks/MakeVariabilityGraphic_20180329-Copy1ForICMETalk.ipynb
\end{figure*}

This proposal relies on a key insight originally from \cite{1986PASP...98..802H} and further developed in \cite{mine}: star trail images are a conduit to achieving subsecond photometry of stellar sources. In star trail images, the tracking is turned off so the telescope rotates with the Earth during the exposure. Stellar sources are stretched into coherent linear trails, which show how the flux of the sources changes throughout the exposure. Figure \ref{fig:trail} shows a simulated LSST star trail image with a one second exposure time. We then train a deep neural network to scan these large, unorthodox images and detect variability (Figure \ref{fig:pipeline}). We assess the performance of our technique on visits and corresponding images that the network was not trained on (Figure \ref{fig:longlimit} shows the results for 15 second trails). The results are competitive with the state of the art \citep{2016SPIE.9908E..0YD}. Below we elaborate on four specific science cases for LSST.

% The input to the network is a 80x80 pixel crop of an LSST star trail image, the output is a binary classification which determines whether the sample are worthy of more detailed, science specific examination. As in many deep learning applications, high quality data and training feedback are essential. We use a suite of LSST simulation tools to produce realistic images. We sample visits from the \textit{minion\_1016} OpSim observing run, then we use CatSim to procure catalogs for each visit, then we use PhoSim to produce high fidelity simulated images of the catalogs \citep{2014SPIE.9150E..15D,2014SPIE.9150E..14C,2015ApJS..218...14P}. We add a new interface into the PhoSim code to simulate \textit{bursts} - a tophat change in flux added to an otherwise flat and static light curve of a source - parameterized by the magnitude change and duration. We train the network over 5 epochs of 80,000 sample 80x80 pixel crops, half of which contain a burst. We train the network to both predict whether the burst exists and to predict the exact photons resulting from the change in flux. Figure \ref{fig:pipeline} highlights this process. 

{\bf Outer Solar System}: The catalog of Solar System objects that live beyond Neptune is largely incomplete below 10 km in diameter, due to the extreme faintness of the reflected sunlight on these distant objects beyond 50 AU. The distribution of these objects at the kilometer size range is crucial for testing planetary formation and evolution models (e.g. \citealt{Kenyon04}). A promising method for observing small and remote Outer Solar System objects is through their rare ($\sim10^{-3}$ per star per year) and brief ($\sim200$ ms) stellar occultations \citep{2013AJ....146...14Z}. These rare, but highly informative detections require a survey with both a large field of view and high time resolution imaging \citep{Bianco09}. Taking star trail images with the LSST would constrain the size distribution and number density of KBOs and Oort cloud objects, leading to further constraints on Solar System evolution models.

{\bf Fast Radio Bursts}: Fast Radio Bursts – extragalactic radio transients of millisecond duration – are among the most intriguing objects in the transient sky. Despite having a prominent occurrence rate (of order many thousands a day over the full sky for the luminosity limit of the searches performed so far, e.g. \citealt{keane}) their nature remains unsettled. Many tens of models have been proposed, a large variety of which leave open the possibility that an optical emission will come in the form of a short burst of duration comparable to that of the radio burst (milliseconds to tens of milliseconds). The upper limits collected so far for the case of the repeating FRB121102 (\citealt{2017MNRAS.472.2800H}, \citealt{2018MNRAS.481.2479M}) are compatible with the observability by LSST – operated in star trail mode – of an FRB lasting 10s of milliseconds and having an intrinsic optical luminosity of the order of ten times that of the 1--3 Gpc distant host galaxy. Using LSST star trails to identify an optical counterpart and the host galaxy of a FRB would pave the way to a direct determination (via spectroscopy) of the red-shift and hence of the distance of the bursting source, allowing the scientific community to exploit the FRBs for many intriguing and unique measurements in cosmology and fundamental physics (see e.g. \citealt{2018NatAs...2..836M} for a review).

{\bf Blazars}: Blazars are a special class of AGN that exhibit extreme variability in all accessible spectral bands. We know of several where during flares, the optical flux would increase by as much as a factor of 100. However, due to observational limitations, we do not know the shortest time scales of their optical variability, which has implications on the size of the emitting region, and the properties of the relativistic jet. Such a program could identify objects for a possible follow-up, either for optical spectroscopy, or more detailed, dedicated monitoring for rapid variability, and have the potential to provide data for cross-correlation of optical flux variability with that measured in other bands \citep{2012AA...545A..48R}. A dedicated white paper by Claudia M. Raiteri describes a survey which employs star trail images to survey blazars and FRBs.

% {\bf Stellar Flares}: 
{\bf Stellar Flares}: Stellar flares are intense releases of energy over a wide range of the electromagnetic spectrum that originate from the reconnection of magnetic field lines in the outer atmospheres of stars. Most flares are rapid and are characterized by an initial rise and decay lasting a few to tens of minutes. By scanning for flare stars with star trail images we can better understand the relation between flares and the masses, ages, and metallicities of their host stars and clusters \citep{2018arXiv180904510S, 2017ApJ...849...36Y}. 

Star trails provide unique data relevant for many applications. There is also the potential to discover new short duration phenomena. The short duration resolution, which comes at no cost to data storage, complements the main LSST database and other surveys, even in different wavelengths. There are many applications for star trails. 

LSST has an unrivaled etendue which makes it the best instrument for star trail imaging - it simply sees more sources. Its rapid readout enables shorter exposures which are better for star trails as well. \textbf{We propose spending 1\% of the LSST survey taking star trails to add this new temporal dimension to LSST's data treasure.}

% \begin{figure*}[htb!]
% \center
% \includegraphics[scale=0.5]{bbox.jpeg}
% \caption{Either going to be 2s star trail performance limits, or we will just have 15s star trail performance, which is currently \ref{fig:longlimit}.}
% \label{fig:shortlimit}
% % http://localhost:8888/notebooks/Code/Astronomy/exp4/notebooks/MakeVariabilityGraphic_20180329-Copy1ForICMETalk.ipynb
% \end{figure*}

\vspace{.6in}
\newpage
\section{Technical Description}
\label{sec:technical}
% \begin{footnotesize}
% {\it Describe your survey strategy modifications or proposed observations. Please comment on each observing constraint
% below, including the technical motivation behind any constraints. Where relevant, indicate
% if the constraint applies to all requested observations or a specific subset. Please note which 
% constraints are not relevant or important for your science goals.}
% \end{footnotesize}

\subsection{Taking Star Trail Images}
\label{sec:overview}

The key element of our proposal is taking star trail exposures with the telescope tracking turned off, or strategically commanded to produce a similar effect. The rotation of the telescope during the exposure with respect to the field produces trails. The trails allow us to see how sources change throughout an exposure. 

Star trail images require new signal analysis to optimize exposure times. Consider the simple scenario of the signal to noise ratio for a single LSST pixel that a source trails over. Let $A$ be the $100\ \mu m^2$ area of the pixel; $N$ be the flux from the source deposited in the pixel; S be the the flux from the sky background; $R$ be the readout noise; $t_{pixel}$ be the 14 ms time the source spends over a single pixel; $t_{exp}$ be the exposure time. Then the signal to noise ratio for resolving a trail amongst the background is:

$$SNR_{trail} = \frac{N\cdot t_{pixel}}{\sqrt{N\cdot t_{pixel} + S \cdot A \cdot t_{exp} + R^2}}$$

\noindent There are additional signal to noise analyses that depend on the signature we are sifting for: bursts, occultations, or gradual changes in flux. These all share the $t_{exp}^{-1/2}$ dependence, which shows that \textbf{shorter exposures provide stronger signal}. This is because the sky background continues to accumulate in the pixel while the source flux is limited by $t_{pixel}$. The operation of the shutter provides a physical lower bound that constrains the minimal exposure time.

The shutter consists of two sets of flat, sliding plates on opposite sides of the focal plane. During each exposure the plates on one side are drawn back to initiate the exposure and the plates on the other side are pulled over to end the exposure. This double act ensures exposure time uniformity across the focal plane. The minimal exposure time is attained by closing the second set of shutter plates immediately after the opposing set of plates finish opening. The exposure time is one second, but the total shutter operation time is two seconds. Two more seconds are required to read out the CCDs and complete the cycle. Thus the minimum exposure cycle is four seconds with a 25\% duty cycle. While this provides the strongest signal, many applications are better served by a higher duty cycle. We envision three primary modes of operation:

\begin{itemize}
\item \textbf{Short Trail:} Taking one second exposures to optimize for signal power and time resolution.

\item \textbf{Long Trail:} Taking 15 second exposures to optimize for duty cycle.

\item \textbf{Strategic Anti-Tracking:} In theory tracking can not only be turned off to induce rotation, but strategically controlled to extend trails further and achieve even higher time resolution by shortening $t_{pixel}$. 
\end{itemize}

We have simulated both short and long trails with high fidelity simulations. For the long trails, we trained a deep neural network to detect bursts in a range of conditions. The detection efficiency for the long trails is shown in Figure \ref{fig:longlimit}. 

\subsection{Constraints}

\begin{figure*}
\center
\includegraphics[width=0.95\columnwidth]{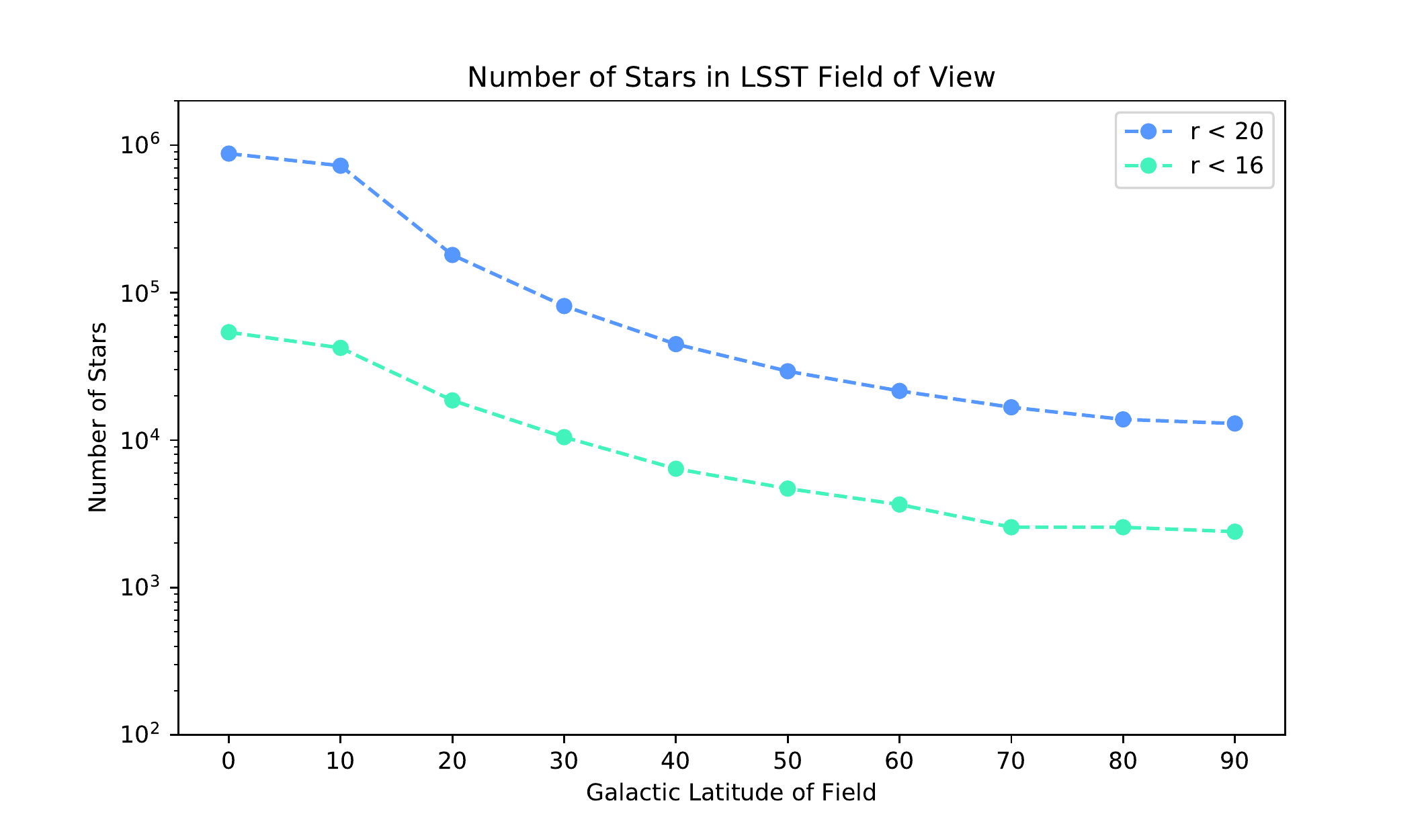}
\caption{The predicted number of stars in the LSST field of view at different galactic latitudes. The counts come from CatSim queries.}
\label{fig:count}
% http://localhost:8888/notebooks/Code/Astronomy/exp8/ProposalFieldCounts.ipynb
\end{figure*}

\begin{itemize}
\item \textbf{Footprint.} There are two modes of footprints. The first mode is to occasionally take short trail images. These observations can be interleaved with the main \textit{Wide-Fast-Deep} survey. It is especially efficient to take short trail images in the middle of a long slew. This shaves 2 seconds off the full cycle time, because the readout can occur while the telescope continues its slewing, and brings the total time cost to the main survey down to 2 seconds. The average LSST slew in the OpSim \textit{minion\_1016} run is 6.5 seconds and 2\% of the simulated observations have a slew time longer than 30 seconds. Inserting a short trail image on just half of these longer slews would meet our science goals and could lead to new discoveries. 

The other mode is a dedicated minisurvey with long trail images. This would be allocated across a small subset of days. This works best for detecting FRBs because we could shadow the LSST with a radio telescope and confirm any discovered optical counterparts.

For both modes it is best to stay in the galactic latitude range $5^o < |b| < 30^o$. The sky brightness washes out the trails for visits too close to the galactic plane and there are fewer sources at higher galactic latitudes (Figure \ref{fig:count}). The length of star trails is related to the declination by $l = 3.75 \cdot \cos(\delta)$ arcminutes per exposure (or $75 \cdot \cos(\delta)$ LSST pixels per second). Visits close to the equatorial plane give the longest trails and best time resolution. 

There are many synergies in taking star trails of fields that have been previously observed with other surveys. The Kepler K2 fields \citep{2014PASP..126..398H} are a good choice because the sources have short cadence lightcurves for many sources and the fields have been extensively studied in other wavelengths as well \citep{2016HEAD...1510605S}.

\item \textbf{Image Quality.} Star trail images are sensitive to the image quality and seeing, measured by the size of the PSF, particularly in the direction transverse to the trails. When the flux of a source is spread out in the transverse direction, the signal decreases. In practice, the performance of the networks we have developed to process star trail images are far more sensitive to the sky background. Thus while good seeing is important, it is not crucial.

\item \textbf{Sky Brightness.} The SNR for resolving trails goes down with $S$, where $S$ is sky brightness (described in Section \ref{sec:overview}). A 4 times brighter sky background reduces the SNR by a factor close to 2, which can dramatically alter the number of trails that can be resolved. Star trail imaging is very sensitive to sky brightness. 

\item \textbf{Total Number of Visits.} Star trail science consists of searching large expanses of sky for rare events. The number of detected events grows linearly with time. Hence the total number of star trail images, or time on the sky, is important.

\item \textbf{Distribution of Visits Over Time and Within a Night.} We would like to shadow the LSST with radio telescopes to synchronize FRB detections on nights where lots of star trail images will be taken. Thus, allocating the star trail visits in a small subset of nights is best. 

\item \textbf{Filter Choice.} We use the \textit{r} filter in our experiments because it has a high transmission efficiency and favorable intersection with the stellar SEDs of interest. Star trail imaging does not exclude the use of other filters that different science cases may require.

\item \textbf{Exposure Constraints.} We described the limits on exposure time in Section \ref{sec:overview}. The primary trade-off is between signal power and duty cycle. The minimal exposure time is one second, which offers the strongest signal and lowest duty cycle. We have also tested 15 second exposures, which have a great duty cycle and weaker signal. 
\end{itemize}

\begin{table}[ht]
    \centering
    \begin{tabular}{|l|l|l|l|}
        \hline
        Properties & Importance \hspace{.3in} \\
        \hline
        Total number of visits & 1\\
        Sky brightness & 1\\
        Distributing visits within a small subset of nights & 2\\
        Filter & 2\\
        Image quality & 2\\
        Individual image depth & 3\\
        Co-added image depth & 3\\
        Number of exposures in a visit & 3\\
        Long-term gaps between visits & 3\\
        \hline
    \end{tabular}
    \caption{Summary of the relative importance of various survey strategy constraints. Each constraint is ranked (1) very important, (2) somewhat important, or (3) not important.}
        \label{tab:obs_constraints}
\end{table}

\section{Performance Evaluation}
% \begin{footnotesize}
% {\it Please describe how to evaluate the performance of a given survey in achieving your desired
% science goals, ideally as a heuristic tied directly to the observing strategy (e.g. number of visits obtained
% within a window of time with a specified set of filters) with a clear link to the resulting effect on science.
% More complex metrics which more directly evaluate science output (e.g. number of eclipsing binaries successfully
% identified as a result of a given survey) are also encouraged, preferably as a secondary metric.
% If possible, provide threshold values for these metrics at which point your proposed science would be unsuccessful 
% and where it reaches an ideal goal, or explain why this is not possible to quantify. While not necessary, 
% if you have already transformed this into a MAF metric, please add a link to the code (or a PR to 
% \href{https://github.com/lsst-nonproject/sims_maf_contrib}{sims\_maf\_contrib}) in addition to the text description. (Limit: 2 pages).}
% \end{footnotesize}

The LSST star trail science use cases follow a common template. They use the LSST's large field to search for rare transient events. These events lead either to follow up observations of individual sources, such as a flaring blazar, or contribute to statistics, such as predicting the number density of small Kuiper Belt objects. Thus, the primary performance metric is a product of the number objects of interest within the field of view times the total time they are exposed in the survey. We sum the weighted contributions from each science in the total performance metric $P$,

$$P = \sum_{s \in Science} t_s w_s \sum_{f \in Field} n_s(f)$$

\noindent where $t_{s}$ is the exposure time for science $s$, $w_s$ is the relative weight of the science application (normalized so that $\sum_{s \in Science} w_s = 1$), and $n_s(f)$ is the number of sources for science $s$ in field $f$.

The relative weights of the aforementioned science applications is a subject of future work. The number of sources $n_s(f)$ depends on the detection limits of our star trail image processing. A simple, practical, and data efficient way these limits can be translated to real images is by injecting a small set of real images with simulated variability. This allows us to estimate the limits that govern $n_s(f)$ with a small amount of data.

\vspace{.6in}

\section{Special Data Processing}
% \begin{footnotesize}
% {\it Describe any data processing requirements beyond the standard LSST Data Management pipelines and how these will be achieved.}
% \end{footnotesize}

We aim to process the data within 48 hours of an exposure. The data will be in the Data Access Center within 24 hours and can be downloaded via the LSST Science Platform. The remaining 24 hours are more than enough to apply the special data processing.

After the instrument signature removal task has been applied, these images must be processed by a custom transient detection pipeline. Our work to date has used neural networks for this task. The network can classify a single 80x80 pixel 1s star trail crop in 0.6ms on a Xeon E5-2640 v4 CPU connected to a NVIDIA Tesla P100 GPU via PCIe. Figure \ref{fig:count} shows that there will be $\mathcal{O}(10^4)$ trails to classify at a typical LSST galactic latitude. Thus, a single GPU setup could process a full LSST image in $\mathcal{O}(30)$ seconds. The crops are processed independently, which allows us to reduce the time required by approximately a factor of $1/M$ where $M$ is the number of GPUs.  

After processing the star trails, we only need to store the the crops and corresponding metadata for detected transient events. We expect detections to be rare enough ($< 1/10,000$ trails) that the required storage will be trivial. Thus a single GPU workstation would provide sufficient computational processing and disk space for the survey we are proposing.

Star trail imaging will also require invoking LSST telescope commands to toggle the tracking. As we continue to validate star trail imaging and do small surveys at other telescopes, we will be implementing and employing similar commands. We do not envision this being a significant burden.

\section{Future Work}
\label{sec:future}

The primary goal of this white paper is to bring star trail imaging and its potential to the attention of the LSST Science Advisory Committee and broader LSST community. Star trail imaging offers a bold new opportunity that the LSST, with its enormous etendue, is uniquely positioned to capture. 

We submitted a proposal for imaging four Kepler K2 fields with star trails with the Dark Energy Camera in the 2019A NOAO observing period. We aim to measure how various stellar variability detection methods, including deep neural networks, translate to real data. We will take this opportunity to gain experience programmatically operating telescopes to induce star trails. Star trail images will also be taken during the LSST commissioning, which will provide further experience and science opportunities.

\section{Acknowledgment}
\label{sec:ack}
This work was developed within the LSST Transients and Variable Stars Science Collaboration (TVS) and the authors acknowledge the support of TVS in the preparation of this paper.

\section{References}

\begingroup
\renewcommand{\section}[2]{}%
\bibliographystyle{aasjournal}
\bibliography{references}
\endgroup

\section*{Author Affiliations}
\parbox{\textwidth}{
\footnotesize
$^{1}$ Institute for Computational and Mathematical Engineering, Stanford University, Stanford, CA, USA\\
$^{2}$ Kavli Institute for Particle Astrophysics and Cosmology, Stanford University, Stanford, CA, USA\\
$^{3}$ Department of Physics, Stanford University, Stanford, CA 94305, USA\\
$^{4}$ Center for Urban Science and Progress, New York University, Brooklyn, NY 11201, USA\\
$^{5}$ Center for Cosmology and Particle Physics, New York University, New York, NY 10003, USA\\
$^{6}$ Department of Astronomy, University of Washington, Seattle, WA 98195-1580, USA\\
$^{7}$ INAF – Osservatorio Astrofisico di Torino, via Osservatorio 20, I-10025, Pino Torinese, Italy\\
$^{8}$ INAF – Osservatorio Astronomico di Cagliari, Loc. Poggio dei Pini, I-09012, Capoterra, Italy\\
$^{9}$ Department of Physics and Astronomy, Purdue University, West Lafayette, IN 47907, USA\\
$^{10}$ Department of Astronomy, University of Illinois at Urbana-Champaign, Urbana, IL 61801, USA\\
$^{11}$ Large Synoptic Survey Telescope Operations, 950 North Cherry
Ave, Tucson, AZ 85719, USA\\
$^{12}$ Lowell Observatory, Flagstaff, AZ 86001, USA\\
$^{13}$ NASA Ames Research Center, Mountain View, CA, USA\\
}
\end{document}